\documentclass[%
 reprint,
superscriptaddress,
 amsmath,amssymb,
 aps,
]{revtex4-2}

\usepackage{graphicx}
\usepackage{dcolumn}
\usepackage{bm}
\usepackage{xcolor}

\begin{document}

\preprint{APS/123-QED}

\title{Magnetocrystalline anisotropy of FeNi and FeCo along the Bain path}

\author{Nica Jane B. Ferrer}
\email{ferrer.ni@northeastern.edu}
\affiliation{%
Department of Physics, Northeastern University, Boston, MA, USA 02115
}%
\author{Gregory A. Fiete}%
\affiliation{%
Department of Physics, Northeastern University, Boston, MA, USA 02115
}%
\affiliation{Quantum Materials and Sensing Institute, Northeastern University, Burlington, MA, 01803 USA}

\date{\today}

\begin{abstract}
We theoretically investigate magnetic anisotropy in materials with non-critical elements to determine which symmetry conditions and atomic shell filling favor enhanced magnetic anisotropy.  We study the magnetocrystalline anisotropies (MCA) of the equiatomic ferrous compounds FeCo and FeNi using \textit{ab initio} calculations  and analytical approaches via the diatomic pair model. We find that when these materials undergo a Bain transformation, that is, the variation of the $a$ and $c$ lattice parameters adjust to interpolate between the B2 and L$1_0$ structural phases while keeping the unit cell volume constant, the MCA versus $r = c/a$ ratio varies differently for FeCo and FeNi despite Co and Ni differing only by one valence electron. To uncover the physics governing these trends, we use a diatomic pair model to perform a theoretical analysis of the \textit{ab initio} results. We find that the MCA variation along the Bain path is correlated with the structural phase of the material as well as the occupation of $(l,m)$-resolved states for each equiatomic ferrous compound. Accordingly, the MCA was found to differ depending on the element paired with Fe to form the Fe-X compound (X = Co, Ni).  Our work
could help guide the scientific community in solving the supply crisis of hard/strong permanent magnets that are crucial for various technological applications such as those depending on motors and generators for energy conversion and clean energy applications. 
\end{abstract}

\maketitle


\section{\label{sec:Introduction} Introduction}


Magnetic materials play an important role in realizing technological systems for various purposes, such as the generation of energy from wind turbines and generators used in modern electric vehicles. The magnetic materials used in these applications are often classified based on how easy or hard it is to magnetize and demagnetize them\cite{Skomski2016} -- a distinction which easily relates the material properties to their potential applications. Such magnets are referred to as either ``soft" or ``hard" depending on how easy or difficult it is to magnetize and demagnetize them. Soft magnets are commonly used for transformers \cite{Silveyra2018} and hard magnets for electric motors and generators \cite{Gutfleisch2011}, like those in electric vehicles and wind turbines. 

This work focuses on hard magnets, also referred to as permanent magnets given their resistance to magnetization and demagnetization. 
\begin{table}
\centering
    \begin{tabular}{c|c}
    \hline
        Material  &  $K_1$ (MJ/m$^3$) \\
    \hline
        Nd$_2$Fe$_{14}$B & 4.9  \\
        Sm$_2$Fe$_{17}$N$_3$ & 8.9   \\
        Sm$_2$Co$_{17}$ & 4.2 \\
        Sm Co$_5$ & 17.0 \\
        Alnico 5,8,9 & 0.32, 0.24, 0.26\\
        L1$_0$ FeNi & 1.0-1.3 \\
        L1$_0$ FePd & 1.71 \\
        L1$_0$ FePt & 15.70 \\
    \end{tabular}
    \caption{Permanent magnets and their magnetocrystalline anisotropy constants $K_1$ (see Refs. \cite{Skomski2016},\cite{Lewis2014, Werwiński_2017, Poirier2015,TANAKA2001118,MARCINIAK2022169347}).}\label{wrap-tab:perm_mag_K1}
\end{table} 
Table \ref{wrap-tab:perm_mag_K1} lists some of the permanent magnets in high demand and their corresponding magnetocrystalline anisotropy (MCA) constants $K_1$, with the MCA being an intrinsic magnetic property related to the dependence of the system energy on the magnetization direction.
The MCA energies shown have been obtained from experimental measurements as well as computational approaches \cite{Skomski2016,Lewis2014,Werwiński_2017,Poirier2015}. Among the major permanent magnets \cite{Gutfleisch2011,Coey2020} used globally are: (i) Nd-Fe-B (more than 60\% of market share \cite{Gutfleisch2011}), (ii) Sm-Co, and (iii) Alnico magnets. It can be seen that Nd-Fe-B and Sm-Co magnets, which dominate the list, involve rare-earth elements like neodymium and samarium.  Such elements are costly and environmentally damaging to mine, separate, and purify \cite{Bourzac2011rare}.  

Given the crucial reliance on these rare-earth permanent magnets, it is not surprising that  there was a rare-earth crisis in 2011 \cite{Coey2020} which points to the possibility of the global demand exceeding the supply in the years ahead, with significant economic consequences. However, by looking at how the magnetocrystalline anisotropy constants $K_1$ of L1$_0$ materials like FeNi, FePd and FePt compare to those of the rare-earth permanent magnets, one is motivated to investigate and identify magnetic materials with mostly non-critical elements that will serve as acceptable substitutes for rare-earth based materials. Our work aims to help guide the scientific community in solving the supply chain crisis of strong permanent magnets in addition to promoting more environmentally friendly processing of magnetic materials for applications by suggesting alternative key elements from the periodic table and the optimal lattice structures of materials that include them. 

Among the materials of interest in this study are equiatomic ferrous compounds like the L1$_0$ materials and other rare-earth lean magnets. By looking at the interplay of crystal symmetry, atomic shell filling, and spin-orbit coupling on the resulting magnetocrystalline anisotropy of rare-earth lean materials, the factors that yield the large magnetocrystalline anisotropy can be determined and used to help select rare-earth lean materials for hard magnets.

The remainder of this paper is organized as follows.  In Sec.~\ref{sec:L10} we review the L1$_0$ crystal structure and describe the Bain path of crystal deformation.  In Sec~\ref{sec:Methodology} we describe the computational and analytical theoretical methods we use to carry out our study.  In Sec.~\ref{sec:Abinitioresults} we present the first-principles {\it ab initio} results of our study, and in Sec.~\ref{sec:Theoreticalanalysis} we present an anlaytical model that captures the dominant trends of the {\it ab initio} study.  Finally, in Sec~\ref{sec:Conclusions} we present the main conclusions of our work and discuss important directions for future study.

\section{L1$_0$ crystal structure and the Bain path}
\label{sec:L10}

\begin{figure}[h!]
{  \includegraphics[width=4cm]{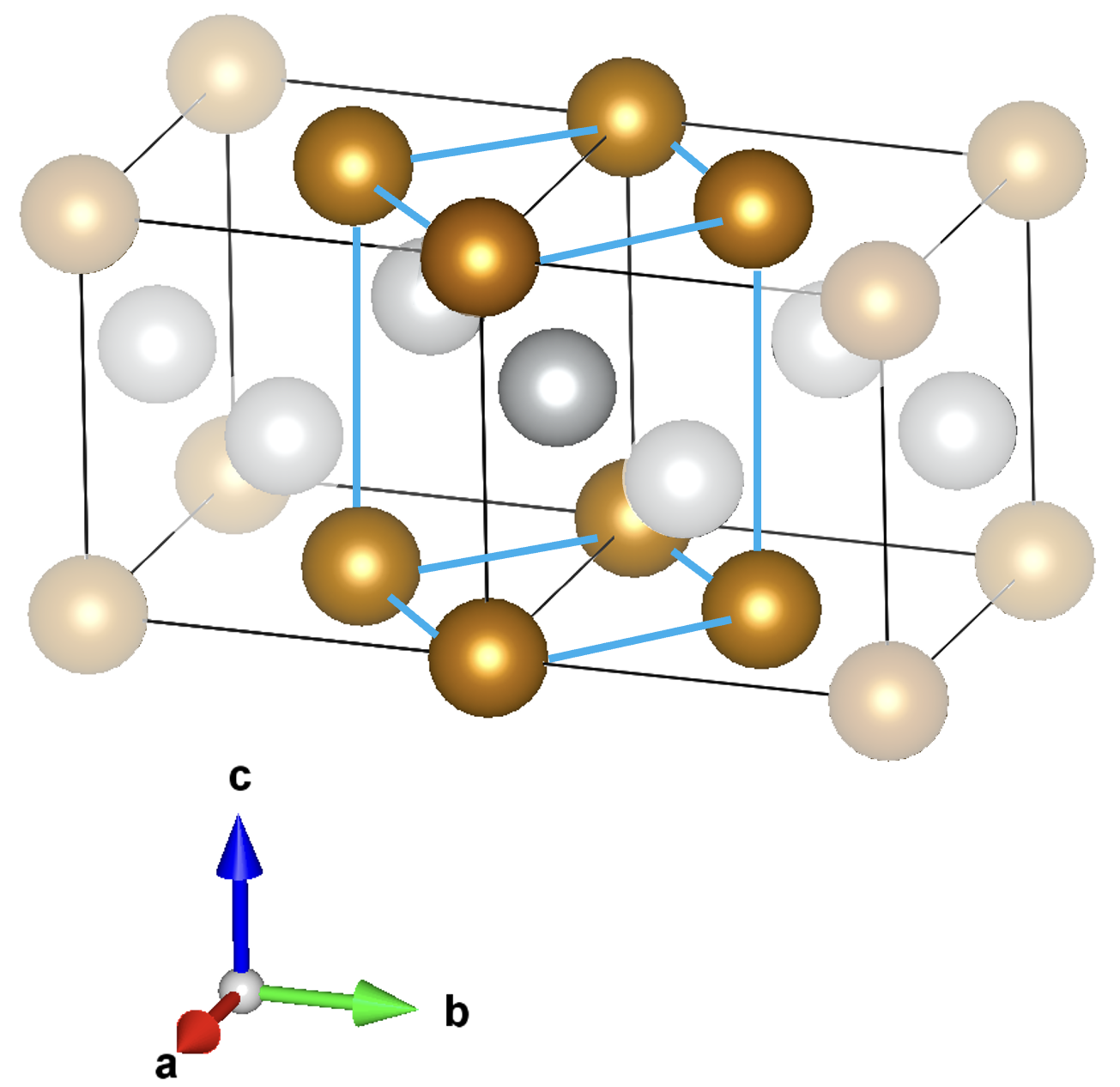}}
\quad
{  \includegraphics[width=4cm]{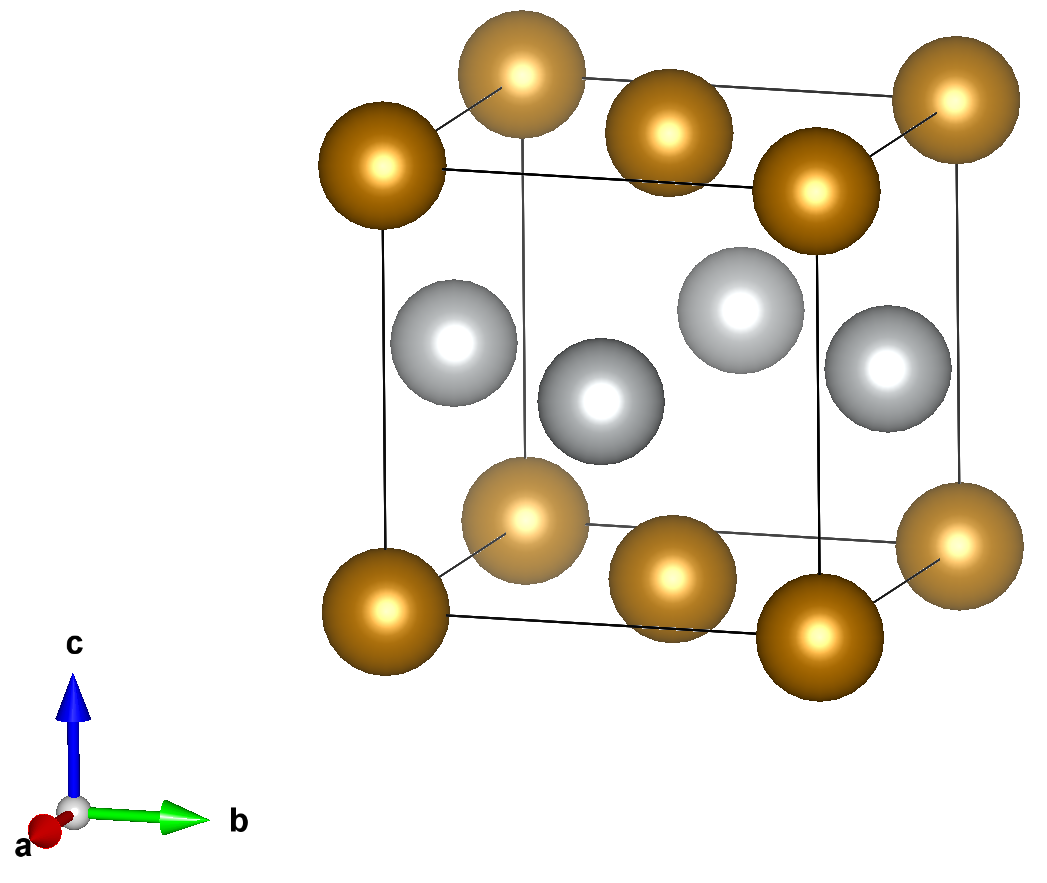}}
\caption{The $tP2$ (left) and $tP4$ (right) unit cells for the L1$_0$ crystal structure \cite{Momma2011_VESTA}.  The coordinate axis indicate the $a,b$ and $c$-axis directions.  The structure is face-centered cubic type of A and B atoms, labelled in different colors.}
\label{fig:tP2_tP4}
\end{figure}

The L1$_0$ crystal structure was first introduced to describe the CuAu structure \cite{Hermann1931, Laughlin2005}. Referring to the structure on the right of Fig.~\ref{fig:tP2_tP4}, the L1$_0$ crystal structure consists of an alternate stacking along the $[001]$ direction of the A and B atoms, which occupy face-centered cubic sites.

As seen in Fig.~\ref{fig:tP2_tP4}, there are two unit cells associated with the L1$_0$ crystal structure: $tP2$ and $tP4$, where in the Pearson notation \cite{Pearson1967} $t$ stands for for the tetragonal crystal system, $P$ for the primitive Bravais lattice, followed by the number of atoms in unit cell (either 2 for $tP2$  or 4 for $tP4$). In our \textit{ab initio} calculations, the $tP2$ unit cell is used.  In our computations we will compute the magnetocrysalline anisotropy as a function of the ratio of $r=c/a$ going from $r=1$ to $r=\sqrt{2}$ while keeping the unit cell volume constant. This path from $r=1$ to $r=\sqrt{2}$ is called the Bain path. A list of different names of structural phases along the path is given in Table~\ref{tab:c_over_a_ratios}.

In our calculations and discussions, we use the symbol I to refer to an intermediate phase (with $1<r<\sqrt{2}$) and \# from 1 to 8 as we subdivide the $r = 1$ to $r = \sqrt{2}$ $c/a$ ratios into 10 $r = c/a$ points along the Bain path. Table~\ref{tab:c_over_a_ratios} summarizes the Bain path for different $r = c/a$ ratios.
\begin{table}[h!]
    \centering
    \begin{tabular}{c|c}
    Phase & $r = c/a$ ratio\\
    \hline
    B2     & 1 \\
    I1     & 1.04602373\\
    I2      & 1.09204746 \\
    I3      & 1.13807119\\
    I4      & 1.18409492\\
    I5      &   1.23011865\\
    I6      &   1.27614237\\
    I7      &   1.3221661 \\
    I8      &   1.36818983\\
    L1$_0$  &  1.41421356
    \end{tabular}
    \caption{The $r = c/a$ ratios along the Bain path which starts from the B2 structural phase ($r = c/a = 1$ ), goes through the I\# intermediate phases (\# =1,2,..,8)  and finally ends at L1$_0$ phase ($r = c/a = \sqrt{2}$)}
    \label{tab:c_over_a_ratios}
\end{table}

\section{\label{sec:Methodology} Methodology}
In this section we describe the computational and analytical theoretical methods we use to carry out our study.

\subsection{\textit{Ab initio} magnetocrystalline anisotropy calculations}

Density Functional Theory (DFT) enables the solving of a many-body problem by reformulating it as an effective one-electron problem with the many-body effects accounted for in the effective potential \cite{Martin_2004}. The Hohenberg-Kohn theorem \cite{HohenbergKohn1964} establishes the existence of a unique electron density corresponding to the external potential $V_{ext}$. The effective one-electron problem is then written in terms of Kohn-Sham orbitals as an \textit{ansatz} \cite{KohnSham1965}. To solve the Kohn-Sham equations and obtain the electron density, various numerical implementations where basis sets or a multiple-scattering theory/Green's function method are used. In this work, the Korringa-Kohn-Rostocker formalism-based DFT implementation \cite{KORRINGA1947392,Kohn1954} via hutsepot \cite{Hoffman2020} was used.

To determine the magnetocrystalline anisotropy of the equiatomic ferrous compounds of interest, an implementation of the torque method \cite{Wang1996} called  MARMOT \cite{Patrick2022} was used. The MARMOT implementation is also able to perform calculations at finite temperature by using the the Disordered Local Moment (DLM) theory \cite{Gyorffy_1985}. DLM theory enables the calculation of magnetic properties at finite temperature by approximating the original statistical physics problem of magnetic moments by an auxiliary problem formulated in terms of local magnetic moments and Weiss fields \cite{Patrick2022}. In this work, the calculations made are at $T = 0$ K temperature.

\subsection{Diatomic-pair model}

In addition to implementing first principles calculations for the Fe monolayer and determining the magnetocrystalline anisotropy of this material, Wang {\it et al.} \cite{Wang1993} also presented the diatomic-pair model which was used to calculate spin-orbit coupling (SOC) energies of diatomic pairs. The justification for this model comes from a phenomenological statistical theory of magnetic annealing demonstrating the importance of diatomic-pairs in predicting magnetic properties of transition metal alloys \cite{JCSlonczewski}.

The diatomic-pair model considers two atoms with partially filled $d$-shells separated by distance $a$. As known from solid state physics, hybridization of $d$-orbitals occurs as each atom experiences the presence of another atom nearby. The orbital angular momentum and $z$-projection of the orbital angular momentum, $(l,m)$, occupations are then determined from the density-of-states (DOS) by noting which states lie below and above the Fermi level. From this energy diagram, the spin-orbit coupling energies for the diatomic pair can be calculated using \cite{Wang1993},
\begin{equation}
    E^{sl} (\boldsymbol{\sigma}) = -\frac{1}{2} \xi^2 \sum_{o,u} \frac{|\langle o|\boldsymbol{\sigma}\cdot \mathbf{L}|u\rangle|^2}{\epsilon_u - \epsilon_o},
    \label{eq:SOC}
\end{equation}
where $\boldsymbol{\sigma}$ is the spin direction, $\mathbf{L}$ is the angular momentum operator, $\xi$ is an overall energy scale of the spin-orbit coupling, and $o, u$ are the occupied and unoccupied states, respectively.

\begin{figure}[h!]
{  \includegraphics[width=7cm]{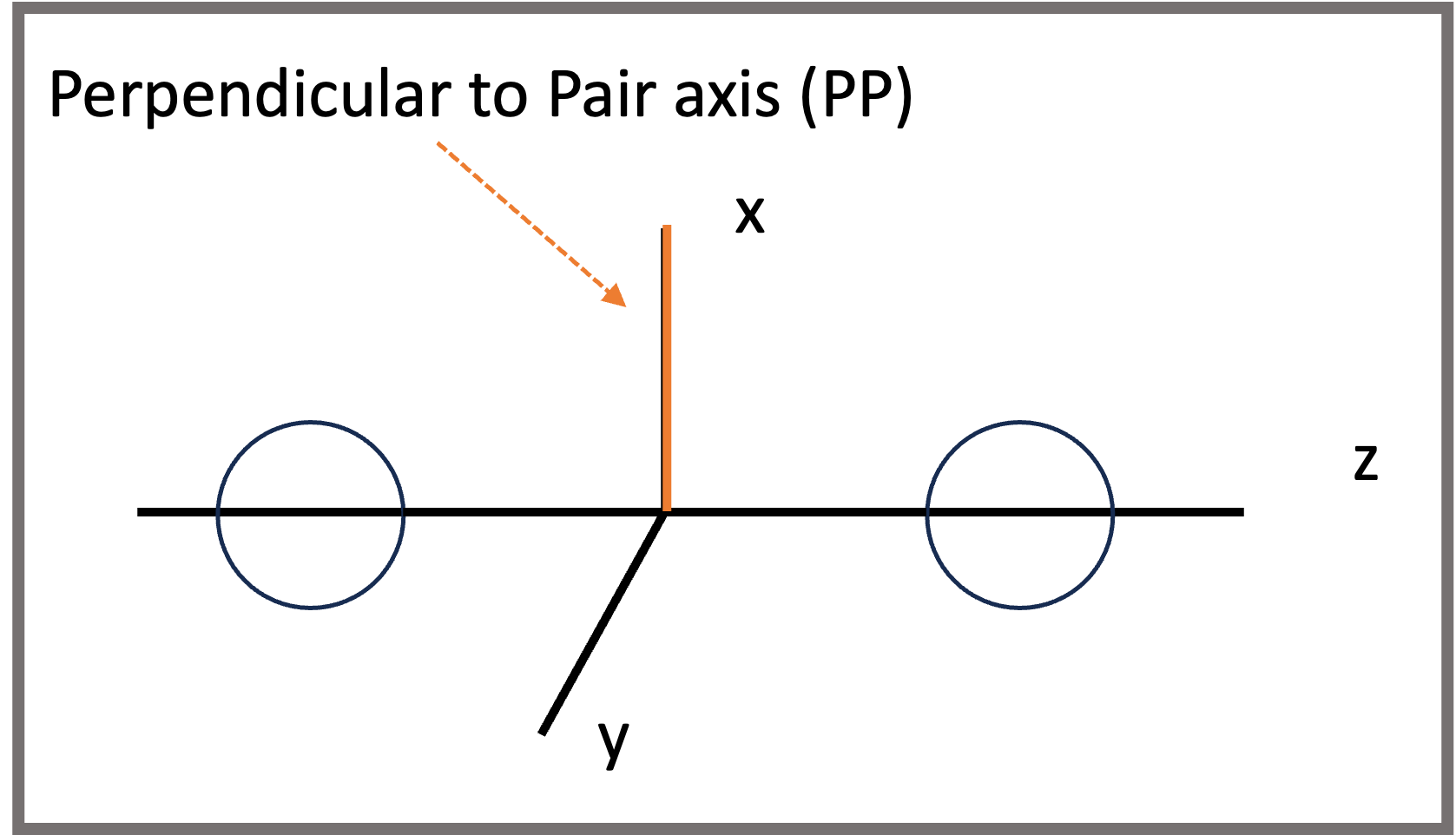}}
\quad
{  \includegraphics[width=7.2cm]{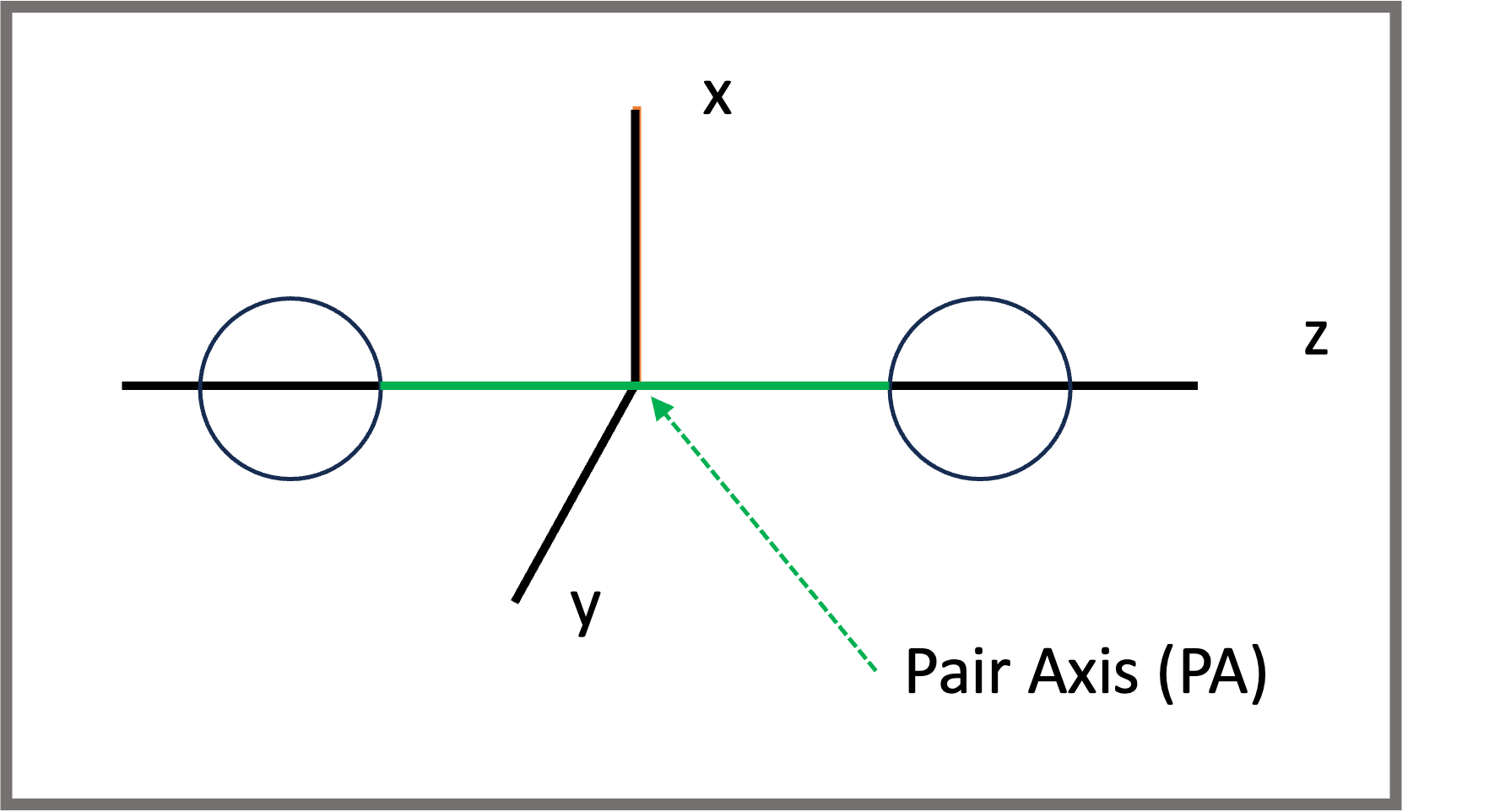}}
\caption{Illustration of important directions relative to the axis joining two atoms in a crystal--Perpendicular to Pair axis (PP) and Pair Axis (PA) directions in the diatomic pair model.  We use these two directions to estimate the magnetocrystalline anisotropy. Figure adapted from Ref. \cite{Wang1993}.}
\label{fig:dia}
\end{figure}

The two main directions of interest when using this model are the PA - Pair Axis direction and PP - Perpendicular to Pair axis direction, as illustrated in Fig. \ref{fig:dia}.

By calculating the spin-orbit coupling energy difference using Eq.\eqref{eq:SOC} between the PA and PP spin orientations, one can estimate the magnetocrystalline anisotropy of a diatomic pair of interest.

\section{\label{sec:Abinitioresults} \textit{Ab initio} results}

\begin{figure}[h!]
{  \includegraphics[width=8cm]{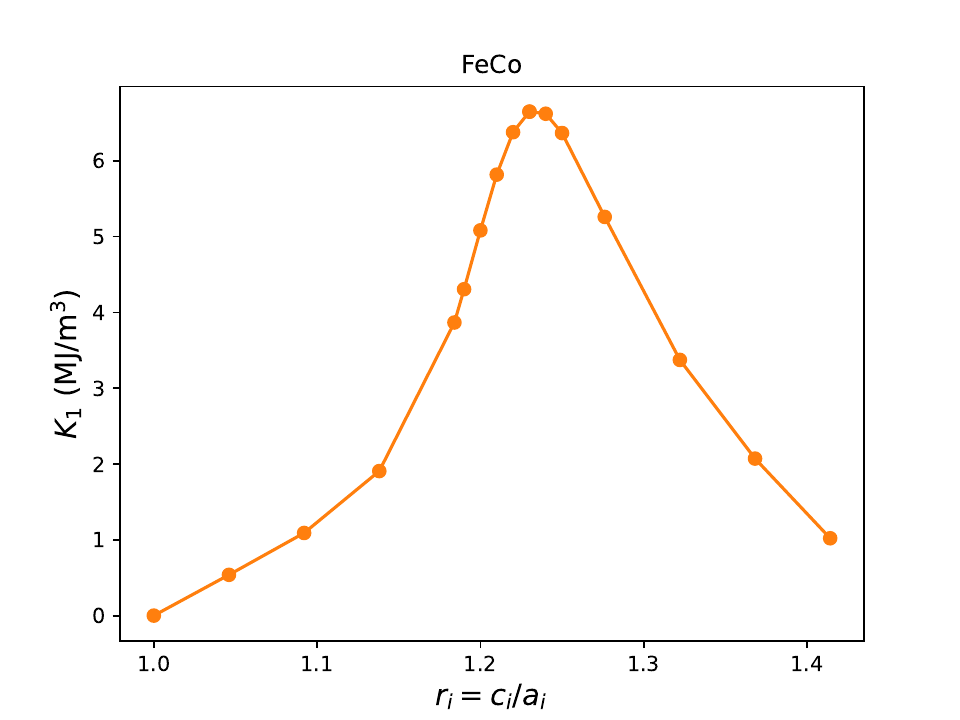}}
\quad
{  \includegraphics[width=8cm]{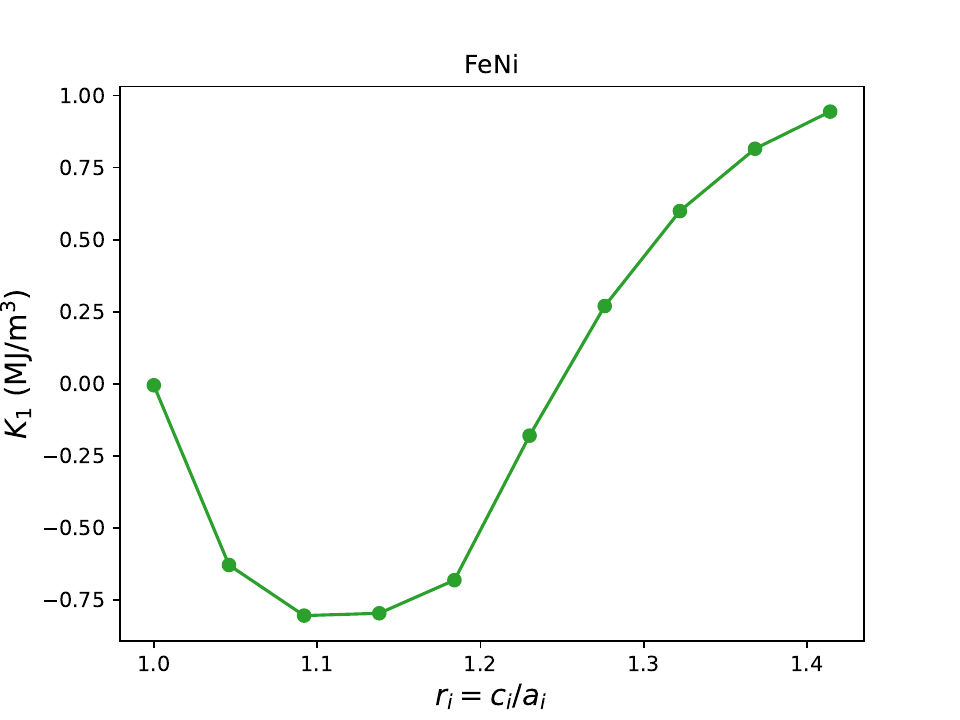}}
\caption{Magnetocrystalline anisotropy constant $K_1$ versus the ratio of lattice parameters $r_i = c_i/a_i$ for FeCo and FeNi. FeNi is more energetically stable in the L1$_0$ phase and FeCo in the B2 phase. It may be possible to combine these materials (by alloying) to  produce a hard magnet with a stronger anisotropy than either one individually.}
\label{fig:K1}
\end{figure}

Figure \ref{fig:K1} shows the magnetocrystalline anisotropy constant $K_1$ (where $K_1$ is the coefficient of a term $K_1\sin^2(\theta)$ in the total energy of the system, where $\theta$ is the angle the total magnetization makes with the easy-axis) as a function of the $c/a$ ratio along the Bain path, described in Sec.~\ref{sec:L10}. 
For FeCo, the $K_1$ versus $r_i = c_i/a_i$ curve increases until it reaches a peak around $r_i \approx 1.23$ and then decreases as the FeCo phase reaches the L1$_0$ point. The order of magnitude of $K_1$ increases as equiatomic element with Fe increases in atomic number $Z$ (moving down the periodic table, not shown). By contrast, the behavior of $K_1$ along the Bain path for FeNi is rather different, and it even changes sign.  However, the trend of increasing $K_1$ as one moves down the periodic table (not shown) is the same as for FeCo. 

Since FeNi is more energetically stable in the L1$_0$ phase and FeCo in the B2 phase, and given the potential of tetragonal FeCo of being an ideal hard magnet, it would be desirable to combine these materials to produce a hard magnet with a stronger anisotropy than either one individually. Based on the trends shown in Fig. \ref{fig:K1}, there is a potential mixture of FeNi and FeCo along the Bain path where is an enhancement and an optimal K$_1$ value achieved. This conclusion is consistent with the results of a recent study on L1$_0$ FeNi$_{1-x}$Co$_x$ alloy in which magnetocrystalline anisotropy is shown to be controlled by the amount of doping \cite{Wysocki2019}. 

\section{\label{sec:Theoreticalanalysis} Theoretical analysis of the \textit{ab initio} results and the diatomic-pair model}
In this section we describe how the \textit{ab initio} results of the previous section can be understood using an analytical model of orbitally-dependent electron hopping between two sites on a lattice.

\subsection{Analysis of the minority channel $(l,m)$-resolved density of states to guide diatomic-pair modeling}
\label{subsec:minority_channel}

\begin{figure*}[h!]
{  \includegraphics[width=8cm]{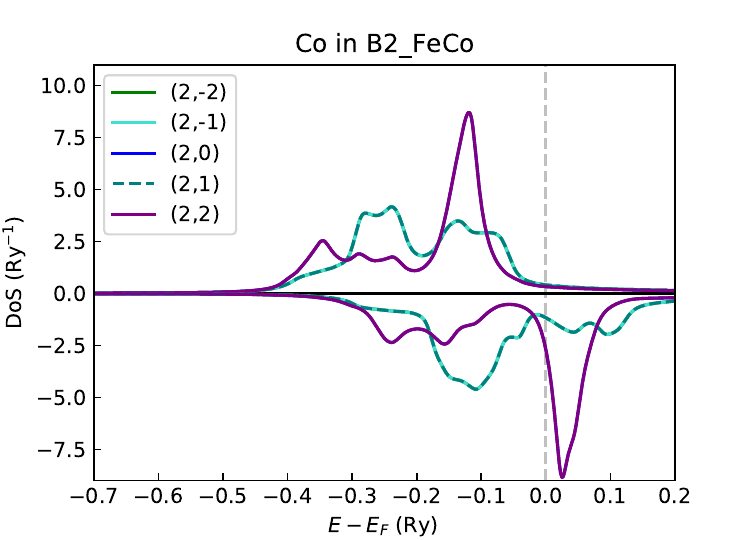}}
\quad
{  \includegraphics[width=8cm]{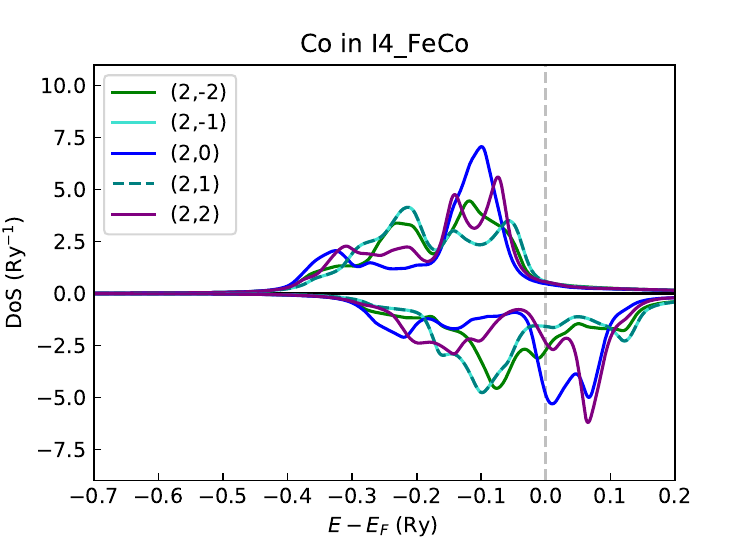}}

{  \includegraphics[width=8cm]{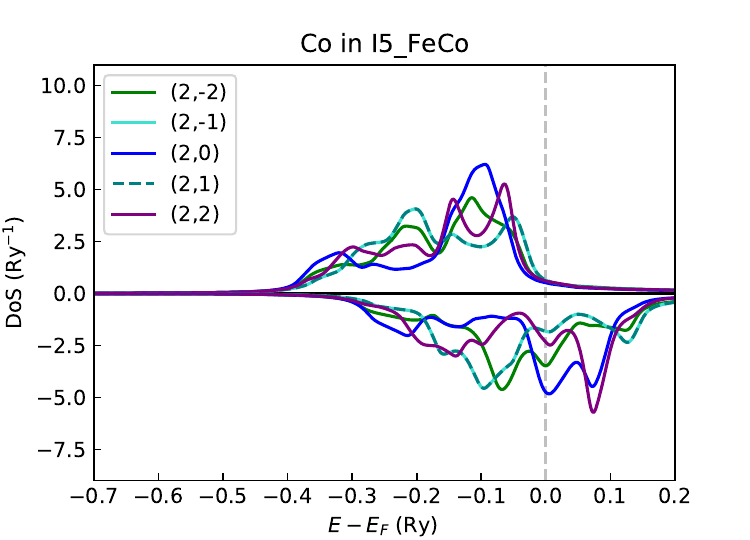}}
\quad
{  \includegraphics[width=8cm]{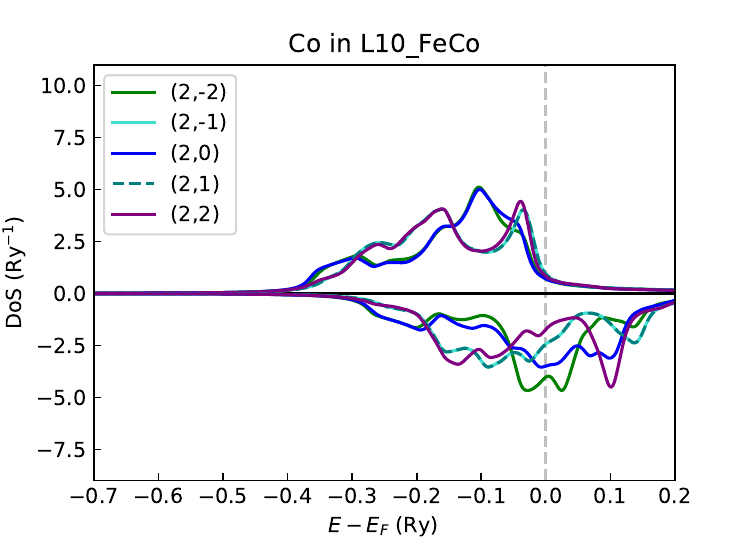}}
\caption{Angular momentum resolved density of states for Co in FeCo. The labeling of the structural phases is given in Table~\ref{tab:c_over_a_ratios}.}
\label{fig:Co_in_FeCo_doslm}
\end{figure*}

\begin{figure*}[h!]
{  \includegraphics[width=8cm]{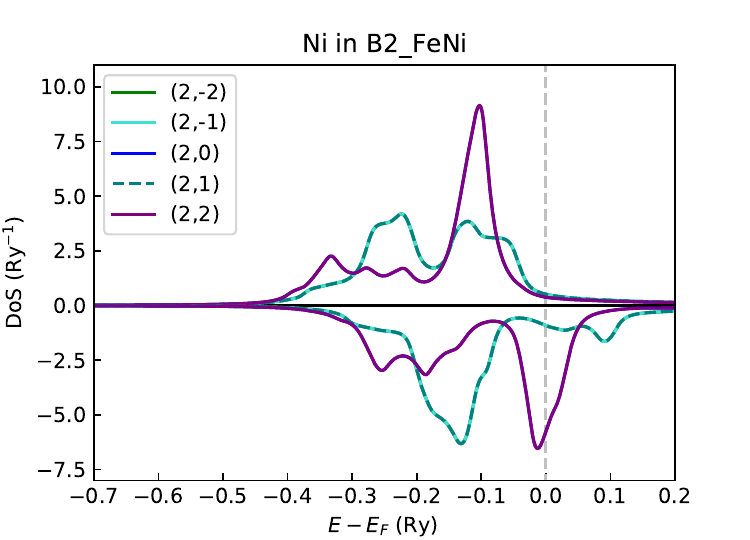}}
\quad
{  \includegraphics[width=8cm]{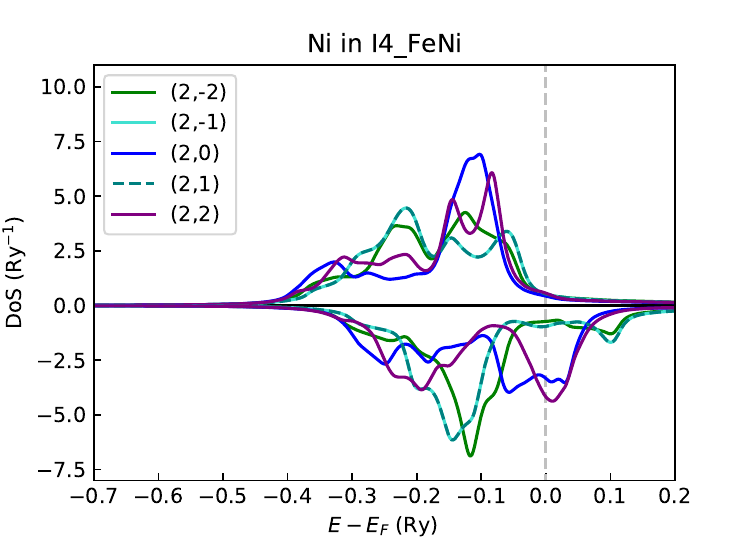}}

{  \includegraphics[width=8cm]{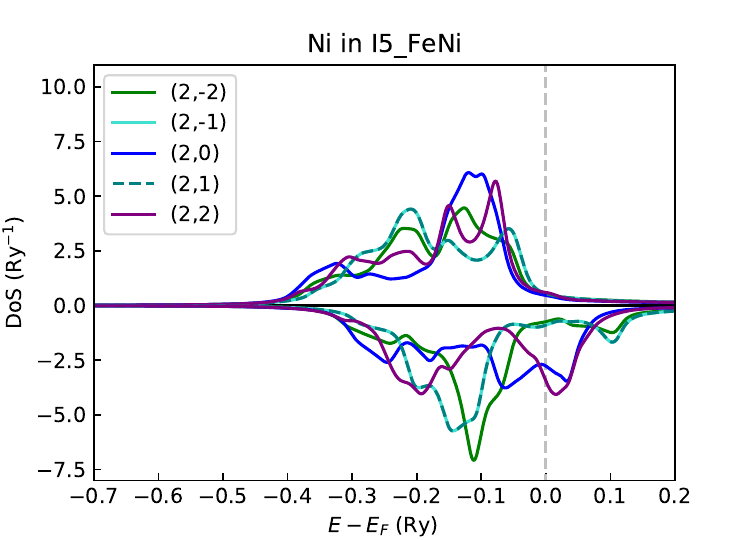}}
\quad
{  \includegraphics[width=8cm]{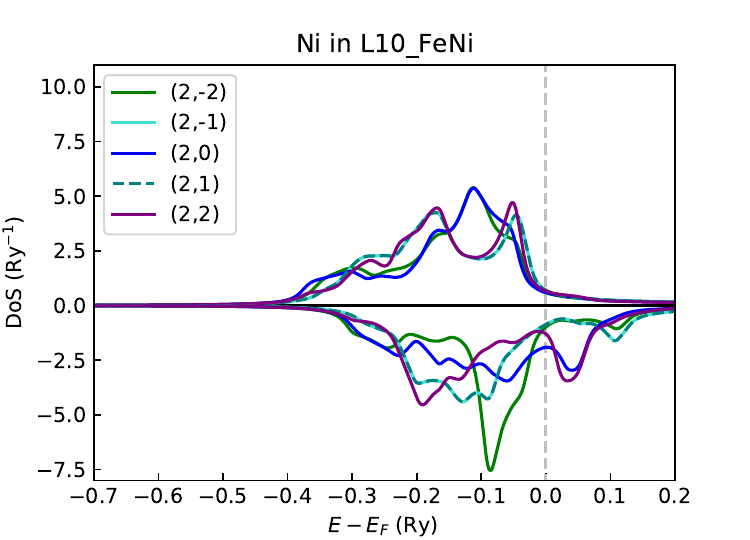}}
\caption{Angular momentum resolved density of states for Ni in FeNi. The labeling of the structural phases is given in Table~\ref{tab:c_over_a_ratios}.}
\label{fig:Ni_in_FeNi_doslm}
\end{figure*}

In order to model the \textit{ab initio} results using a diatomic-pair model, we consider the behavior of the angular momentum $(l,m)$-resolved density of states along the Bain path. 
Figures \ref{fig:Co_in_FeCo_doslm}, \ref{fig:Ni_in_FeNi_doslm} show the angular momentum $(l,m)$-resolved density of states contributions from Co and Ni to FeCo and FeNi, respectively. It can be seen that there are strong overlaps among various $(l=2, m)$ states in the B2 phase for both materials, which are indicative of degeneracies in the density of states curves.

\begin{table}[h!]
    \centering
    \begin{tabular}{c|c}
        Curve & Angular momentum $(l,m)$-resolved states ($\downarrow$)\\
        \hline
         Green &  $(2,-2)$, $(2,-1)$, $(2,1)$\\
         Purple & $(2,0)$ $(2,2)$
    \end{tabular}
    \caption{Corresponding degeneracies in the $(l,m)$-resolved density of states in the B2 phase for Ni and Co in FeNi and FeCo, respectively (Figs. \ref{fig:Co_in_FeCo_doslm}, \ref{fig:Ni_in_FeNi_doslm}). }
    \label{tab:degeneracies_B2_DOS}
\end{table}

In summary, the two main degeneracies are: (i) $m = -2, -1, 1$ and (ii) $m = 0,2 $. While crystal field effects are negligible in transition metal alloy magnets \cite{Skomski2010_JAppPhy}, it is interesting to see the correlation of the B2 phase DOS degeneracies with the state degeneracies of cubically symmetric crystal fields \cite{Skomski2020_HMMM}. 
\begin{table}[h!]
    \centering
    \begin{tabular}{c|c}
        Curve & Angular momentum $(l,m)$-resolved states ($\downarrow$)\\
        \hline
         Dashed green &  $(2,-1)$, $(2,1)$\\
         Purple & $(2,2)$\\
         Blue & $(2,0)$ \\
         Green & $(2,-2)$
    \end{tabular}
    \caption{Corresponding degeneracies in the $(l,m)$-resolved density of states in the L1$_0$ phase for Ni and Co in FeNi and FeCo, respectively (Figs. \ref{fig:Co_in_FeCo_doslm}, \ref{fig:Ni_in_FeNi_doslm}). }
    \label{tab:degeneracies_L10_DOS}
\end{table}
\noindent Among the general features seen in both cases are:
\begin{itemize} 
\item Majority spin channels are filled for all phases from B2 to L1$_0$,
\item Drastic changes occur on how states are filled in the minority spin channel along the Bain path (B2 to L1$_0$),
\item In phase I4 (I: intermediate $r = c/a$ values between B2 and L1$_0$ phase), it can be seen that while the degeneracies are now lifted, the initially overlapping states are still close to one another with respect to energy/$x$-axis shifts and peak changes, 
\item In phase I5, the DOS of each state in the spin down channel have become more distinct from one another, but in phase L1$_0$, it is important to consider the dominant peaks to account for the peak magnitude variations in the I5 to L1$_0$ regime. 
\end{itemize}

Based on these observations, two regimes along the Bain path are identified based on state degeneracies and phase-specific occupations of $(l, m)$ states: Regime I, B2-like and Regime II, L1$_0$-like degeneracy regimes.  We will use this insight to describe how the mangnetocrystalline anisotropy parameter $K_1$ changes with crystal distortion along the Bain path.


\subsection{Approximation of the Slater-Koster parameters for the Fe-Ni/Co diatomic pair }
\label{subsec:pair_model}

Given the simple crystal symmetries of the materials of interest, a two-step approximation for the Slater-Koster parameters based on pairs of sites (Fig.~\ref{fig:dia}) may be employed. First, the transferable tight-binding fit parameters from the work of Shi and Papaconstantopoulos \cite{ShiPapa2004} are used. The interatomic distance between A-A species, $d_{A-A}$ of atoms $\rho_{A}$ is given by,
\begin{equation}
    \rho_{A} = r*d_{A-A} = \frac{c}{a} d_{A-A} \:\: \text{(A: Fe, Co, Ni)},
\end{equation}
where $c$ and $a$ are the lattice constants. 

The tight-binding fit parameters are then given by
\begin{equation}
    V_{d d \gamma}(\rho) = \eta_{d d \gamma} \frac{\hbar^2}{m\rho^{5}},
\end{equation}
where the constants $\eta_{d d \gamma}$ for different elements are tabulated in Ref.\cite{ShiPapa2004}. Given the use of a tP2 unit cell (Fig. \ref{fig:tP2_tP4}), it is easy to see that the underlying crystal lattices A-A and B-B are cubic for both A = Fe and B = Ni, Co.

The desired Slater-Koster parameters for the A-B diatomic pair is then obtained as the geometric average of the A and B parameters \cite{Zemen2014,BallhausenGray},
\begin{equation}
    V^{Fe, X}_{d d \gamma} = \sqrt{V^{Fe}_{d d \gamma} V^{X}_{d d \gamma}}.
\end{equation}.

\subsection{Diatomic-pair model for FeNi and FeCo in \\ (i) B2-like and (ii) L1$_0$ phase regimes}

Having identified the relevant states for the diatomic-pair model from the DOS analysis and given the approximate Slater-Koster parameters of the diatomic pairs in the previous subsection, the anisotropy energy from the spin-orbit coupling energy difference for spin alignments along the pair-axis (PA) and perpendicular to the pair axis (PP) [see Fig.~\ref{fig:dia}] can be determined from,
\begin{equation}
    \Delta E = E^{dd}(PA)-E^{dd}(PP) = E^{dd}(\sigma_z) - E^{dd}(\sigma_x), 
\end{equation}
by making use of Eq.\eqref{eq:SOC}.

\begin{figure*}[h!]
{  \includegraphics[width=16cm]{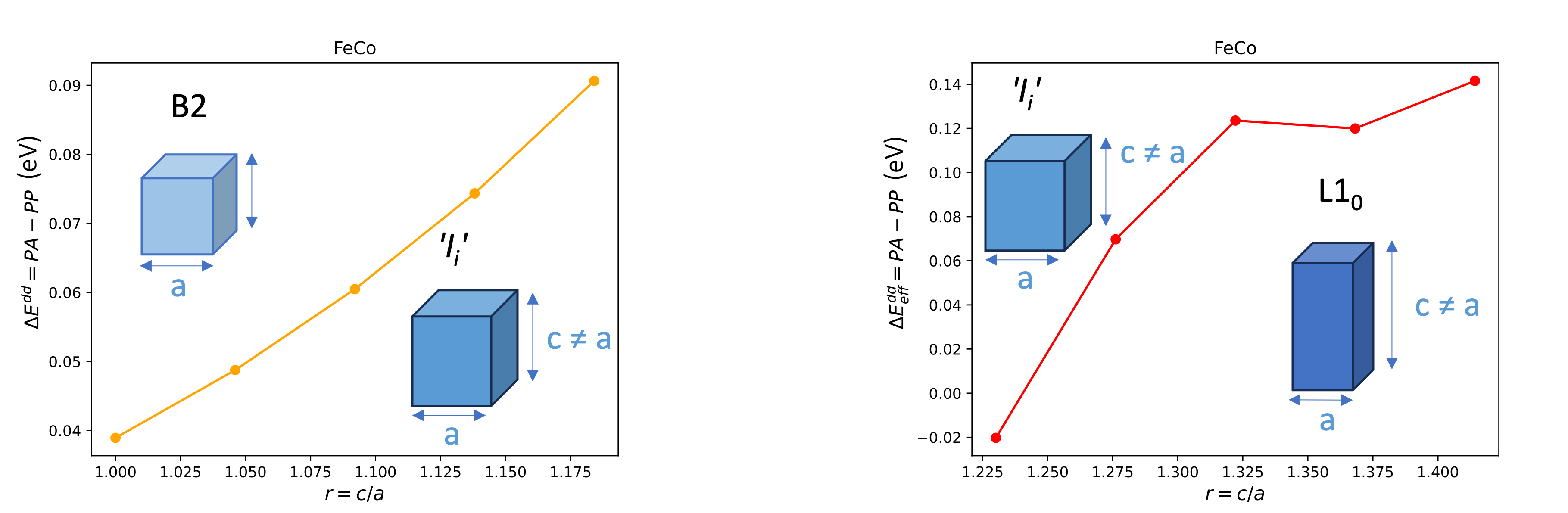}}
\caption{Energy difference PA-PP, $ \Delta E = E^{dd}(PA)-E^{dd}(PP) = E^{dd}(\sigma_z) - E^{dd}(\sigma_x),$ as a function of $r = c/a$ ratio for FeCo computed from the analytical pair model in Sec.~\ref{subsec:pair_model}. The left plot corresponds to phases from B2 to I4, and the right plot to I5 to L1$_0$ structural phases.  See Table~\ref{tab:c_over_a_ratios}. It can be seen that the diatomic pair model prediction agrees trend-wise with the \textit{ab initio} results (See Fig. \ref{fig:K1}) in the B2-like degeneracy regime. FeCo is known to be more energetically stable at the B2 phase according to \textit{ab initio} calculations.} 
\label{fig:FeCo_finalres}
\end{figure*}

Figure \ref{fig:FeCo_finalres} shows the energy $\Delta E$ corresponding to the analytical expressions derived for the diatomic pair as a function of $r = c/a$ ratio along the Bain path for FeCo. The orange and red curves correspond to Regime I (B2-like degeneracies) and Regime II (L$1_0$-like degeneracies), respectively. It can be seen that in Regime I, the diatomic-pair model for FeCo predicts that the anisotropy energy should increase with respect to the $r = c/a$ increase. In Regime II, the model predicts that the anisotropy energy will increase and then decrease around $r = 1.37$, and then finally increases again when in the L$1_0$ structure. Comparing the predictions of the diatomic-pair model for FeCo with the MCA vs the $r = c/a$ ratio trend from the \textit{ab initio} calculations in Regime I -- the regime that is closer to the B2 phase -- one sees a similar trend: FeCo is most energetically favorable in the B2 phase.

\begin{figure*}[h!]
{  \includegraphics[width=16cm]{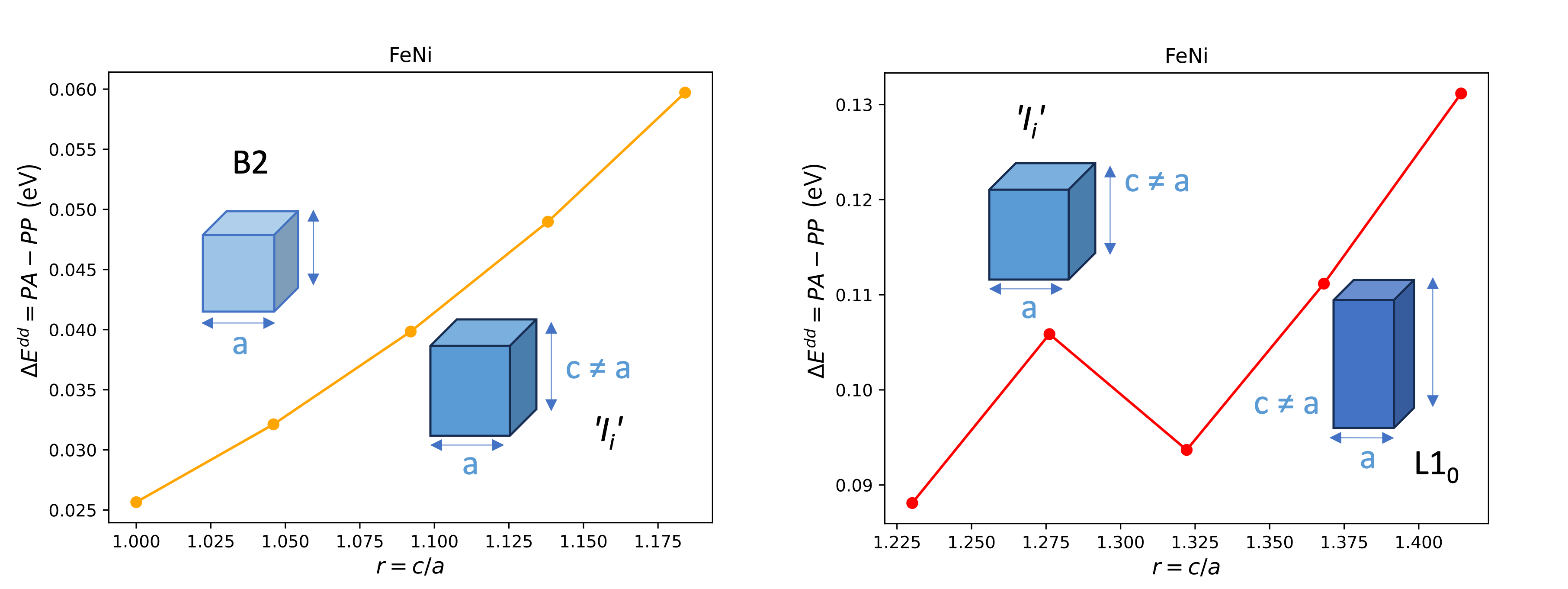}}
\caption{Energy difference PA-PP, $ \Delta E = E^{dd}(PA)-E^{dd}(PP) = E^{dd}(\sigma_z) - E^{dd}(\sigma_x),$ as a function of $r = c/a$ ratio for FeNi computed from the analytical pair model in Sec.~\ref{subsec:pair_model}. The left plot corresponds to phases from B2 to I4, and the right plot to I5 to L1$_0$ structural phases. See Table~\ref{tab:c_over_a_ratios}. It can be seen that the diatomic pair model prediction agrees trend-wise with the \textit{ab initio} results (See Fig. \ref{fig:K1}) in the L1$_0$-like degeneracy regime. FeNi is known to be more energetically stable at the L1$_0$ phase according to \textit{ab initio} calculations.}
\label{fig:FeNi_finalres}
\end{figure*}

For FeNi, as can be seen in Fig. \ref{fig:FeNi_finalres}, the diatomic-pair model predicts an increase in the anisotropy energy in Regime I and in Regime II. The anisotropy energy initially increases from the I5 to I6 phase, but decreases again in I7 phase and then increases again from there on.  See Table~\ref{tab:c_over_a_ratios} for a definition of the I5, I6, and I7 phases.

Interestingly, the diatomic-pair model is in agreement with \textit{ab initio} results for the regime closer to L$1_0$. For FeNi, the energetically favorable phase is the L$1_0$ phase.  In summary, the diatomic-pair model correctly predicts the trends of energy/magnetocrystalline anisotropy/crystal structure for both FeCo and FeNi.

\section{\label{sec:Conclusions} Conclusions}

The variation of the magnetocrystalline anisotropy along the Bain path for equiatomic ferrous compounds was investigated using (i) \textit{ab initio } calculations and (ii) the diatomic-pair model. It was found from \textit{ab initio} calculations that the anisotropy varies in different ways along the Bain path for FeCo and FeNi despite Co and Ni only differing by 1 valence $d$-electron.  This is an important message for guiding the search for hard magnets without rare earth elements.

By performing a qualitative analysis of the atomic species-resolved density of states, it was found that the spin-down states play a crucial role in predicting the magnetic properties of FeCo and FeNi given its drastic variation along the Bain path.  Building from an extended analysis of published results on Fe monolayers, the diatomic-pair model of Wang {\it et al.} \cite{Wang1993} was used for explaining the magnetocrystalline anisotropy variation along the Bain path. The diatomic-pair model for equiatomic ferrous compounds is: (i) material-specific with respect to the structural phase-specific $(l,m)$ occupations of the $(l,m)$-resolved density of states and (ii) implemented differently for the two regimes given the phase transition from B2 to L$1_0$ through intermediate states I\# (\# corresponds to a specific $r = c/a$ ratio along the Bain path; see Table~\ref{tab:c_over_a_ratios} for the definition of this phase notation). 

Along the Bain path, the occupation of each $(l, m)$ state drastically changes and consequently this yields different analytical expressions for the SOC diatomic pair energies given by Eq.~\eqref{eq:SOC}: The difference between diatomic pair energies with magnetization along the Pair Axis (PA) and Perpendicular to the Pair axis (PP).  See Fig.~\ref{fig:dia} for a schematic and Fig.~\ref{fig:FeCo_finalres} and Fig.~\ref{fig:FeNi_finalres} for the numerical trends for FeCo and FeNi.

By comparing the predictions of the diatomic-pair model for FeCo and FeNi to \textit{ab initio} results, the variation of the MCA along the Bain path was found to be correlated with the phase-specific occupation of the $(l, m)$ states. The latter also depends on the element paired with Fe to form an equiatomic ferrous compound Fe-X (X = Co, Ni). 

In summary, the diatomic pair model implemented for the two regimes, depending on the degeneracy lifting, is able to partially capture the magnetocrystalline anisotropy vs $r = c/a$ trend for FeCo (Regime I closer to the B2 phase) and FeNi (Regime II closer to the L1$_0$ phase).  The approach we have used here could be applied to other materials to help provide a guide and rules-of-thumb for understanding how to maximize the hard magnet potential with elemental and crystal structure specificity.  Our method could help address the various environmental and international issues around rare-earth based magnets by redirection attention to rare-earth free or rare-earth lean materials.

\acknowledgements
We acknowledge important discussions with Benjamin J. Weider, Julie B. Staunton, George Marchant, Chris Woodgate, and Laura H. Lewis.  We gratefully acknowledge funding from the Department of Energy under (BES) Award No. DE-SC0022168,


\bibliography{apssamp}

\end{document}